\documentclass[aps,prl,superscriptaddress,reprint]{revtex4-1}
\usepackage{amsmath}
\usepackage{amsfonts}
\usepackage{amssymb}
\usepackage{graphicx}
\usepackage{array}

\begin{document}

\title{Anomalous Gap Reversal of the $3+1/3$ and $3+1/5$ Fractional Quantum Hall States }

\author{Ethan Kleinbaum}
\affiliation{Department of Physics and Astronomy, Purdue University, West Lafayette, IN 47907}
\author{Ashwani Kumar}
\affiliation{Department of Physics, Monmouth College, Monmouth, IL 61462}
\author{L.N. Pfeiffer}
\affiliation{Department of Physics, Princeton University, Princeton, NJ 08544}
\author{K.W. West}
\affiliation{Department of Physics, Princeton University, Princeton, NJ 08544}
\author{G.A. Cs\'athy}
\affiliation{Department of Physics and Astronomy, Purdue University, West Lafayette, IN 47907}
\affiliation{Birck Nanotechnology Center, Purdue University, West Lafayette, IN 47907}

\date{\today}

\begin{abstract}
% The nature of the odd-denominator fractional quantum Hall states developing in the second Landau 
% level is currently unknown.
%  In this work focusing on the upper spin branch of the second Landau level

In this work we report the opening of an energy gap  at the filling factor
$\nu=3+1/3$, firmly establishing the ground state as a fractional quantum Hall state.
This and other odd-denominator states unexpectedly break particle-hole symmetry. 
Specifically, we find that the relative magnitudes of the energy gaps of the $\nu=3+1/3$ and $3+1/5$ states
from the upper spin branch are reversed when compared to the 
$\nu=2+1/3$ and $2+1/5$ counterpart states in the lower spin branch.  
Our findings raise the possibility that the former states have a non-conventional origin.

%question the conventional nature of the $2+1/5$, $3+1/5$ states and
%are suggestive of a much richer physics.

%Specifically, the relative magnitudes of the energy gaps of the $3+1/3$ and $3+1/5$ states
%from the upper spin branch are reversed when compared to the $\nu=2+1/3$ and $2+1/5$ counterpart 
%states in the lower spin branch.  
%Furthermore, at the symmetry-related filling factor $\nu=3+2/3$ we find that
%the ground state in our sample is not a fractional quantum Hall state.
%We explore several possible sources of our observations. 

% We report the opening of an energy gap for a new fractional quantum Hall state at the filling factor
% $\nu=3+1/3$ forming in upper spin branch of the second Landau level.
%This and other odd-denominator states unexpectedly break particle-hole symmetry across
%different spin branches.

%our estimats of the disorder-free energy gap for the $\nu=2+1/5$ state 
%are one order of magnitude less than theoretical predictions.

\end{abstract}
\maketitle

Over the last three decades we have wittnessed an ongoing exploration of topological phenomena in electronic systems.
% The study of topological degree of freedom permeates contemporary condensed matter physics.
Topological ground states may arise from either single-particle band structure effects \cite{iqh,topoinsul}
or from emergent  many-body effects in strongly interacting systems.
%Currently there is an intense effort lavished on many-body systems in which emergent topological properties 
%are due to strong interactions.  
One example of the latter is the  fractional quantum Hall state (FQHS) 
at the Landau level filling factor $\nu=1/3$ \cite{tsui}, a ground state belonging to the larger class 
of conventional Laughlin-Jain FQHSs \cite{Laughlin,Jain}.
%These states are also referred to as Abelian quantum liquids.

More recently it was realized that the family of topological ground states 
may be much richer than previously thought. Of the novel FQHSs the ones 
supporting non-Abelian quasiparticles have generated the most excitement \cite{moore,wen,nayak}.
The FQHS at $\nu=5/2$ is believed to be such a non-Abelian state \cite{willett}. 
However, several other FQHSs in the region  $2<\nu<4$, commonly called the second Landau level (SLL),
are also thought to be non-Abelian 
\cite{ReadRezayi,BondersonSlingerland,Ardonne,Simon,WojsJain,Balram}.

Despite sustained efforts in theory \cite{ReadRezayi,BondersonSlingerland,Ardonne,Simon,WojsJain,Balram}, 
the nature of the prominent odd-denominator FQHSs forming in the SLL,such as the ones at $\nu=2+1/3$ and $2+1/5$,
remains unknown. The FQHSs at $\nu=2+1/3$ 
\cite{pan99,pan08,xia04,dean08,choi08,kumar10,nuebler10}
admits both non-Abelian candidate states
%quantum liquids which, have several non-Abelian candidate states
\cite{ReadRezayi,BondersonSlingerland}
as well as a conventional Laughlin-Jain description \cite{Laughlin,Jain}. 
% Theoretical work on
% wavefunction overlap and on the excitations of the $\nu=2+1/3$ FQHS
% suggest a non-Laughlin, and therefore an exotic ground state [Wijs,Balsam,Peterson]. 
The relatively poor overlap between the exact and numerically obtained
wavefunctions \cite{wojs2,Toke,Papic,PetersonJolicoeur,SimionQuinn,wojs1} and the unusual excitations  \cite{Balram}
does not provide firm evidence for Laughlin correlations in the $\nu=2+1/3$ FQHS.
A number of recent experiments of the $\nu=2+1/3$ FQHS, however, found its bulk  \cite{kumar10} 
and edge  \cite{dolev,willett2,ensslin}  properties consistent with a Laughlin description.
The  other prominent FQHS at $\nu=2+1/5$   \cite{dean08,choi08}
is generally believed to be of the conventional Laughlin type \cite{Papic,PetersonJolicoeur,SimionQuinn,wojs1},
although there is a non-Abelian construction for it as well \cite{BondersonSlingerland}.
%  Experiments on the bulkand the edges of the $\nu=2+1/3$ FQHS so far are 
%  consistent with the Laughlin-Jain picture, while data on the $\nu=2+1/5$ FQHS remains scarce. 
It is therefore currently not clear whether or not the prominent odd-denominator FQHSs in the SLL,
such as the ones at $\nu=2+1/3$ and $2+1/5$, 
require a description beyond the conventional Laughlin-Jain theory.

Experiments on the odd-denominator FQHS in the SLL have been restricted almost exclusively to the $2 < \nu <3$ range, called the 
lower spin branch of the SLL (LSB SLL). Motivated by their poor understanding,  
we have performed transport studies of these FQHSs in the 
little known upper spin branch of the SLL (USB SLL), i.e. in the $3<\nu<4$ region.
%     Our study constitutes an important test of the odd-denominator FQHSs since it allows for a check of the expected
%     particle-hole symmetry relationship of states of different spin branches.
We establish a new FQHS at $\nu=3+1/3$ by detecting the opening of an energy gap.
A quantitative comparison of the gap at this and other filling factors reveals two surprising findings:
%, which seem to break the expected particle-hole symmetry between the different spin branches:
1) the ground state at $\nu=3+2/3$, a symmetry-related filling factor to $\nu=3+1/3$,
is not a FQHS, despite the existence of a strong depression in the longitudinal magnetoresistance and 
2) most intriguingly, the activation energy gaps $\Delta$ of the prominent odd-denominator FQHSs are
reversed across different  spin branches of the SLL. Indeed, in stark contrast to the well
established relation $\Delta_{2+1/3} > \Delta_{2+1/5}$ between the gaps of FQHSs of the
the LSB SLL, in the USB SLL we find $\Delta_{3+1/3} < \Delta_{3+1/5}$.
%                 We explore several possible sources of this anomalous reversal of the gaps.
Within the conventional Laughlin-Jain picture we are unable to account for this anomalous gap
reversal. Our result raises therefore the possibility of a non-conventional origin at least for
a subset of the FQHSs of the upper spin branch.
% and conclude that the $\nu=3+1/5$ FQHS is likely stronger than expected.
%
%Our observations suggest that by controlling the effective electron-electron interaction 
%in the USB SLL one can tune
%the topological ground state of the odd-denominator FQHSs in the SLL.

%These findings suggest that the $\nu=2+1/3$ FQHS may not be well 
%described by the Laughlin-Jain wavefunction 
%and call for the examination of alternative non-Abelian theories. 

\begin{figure*}
 \includegraphics[width=1.95\columnwidth]{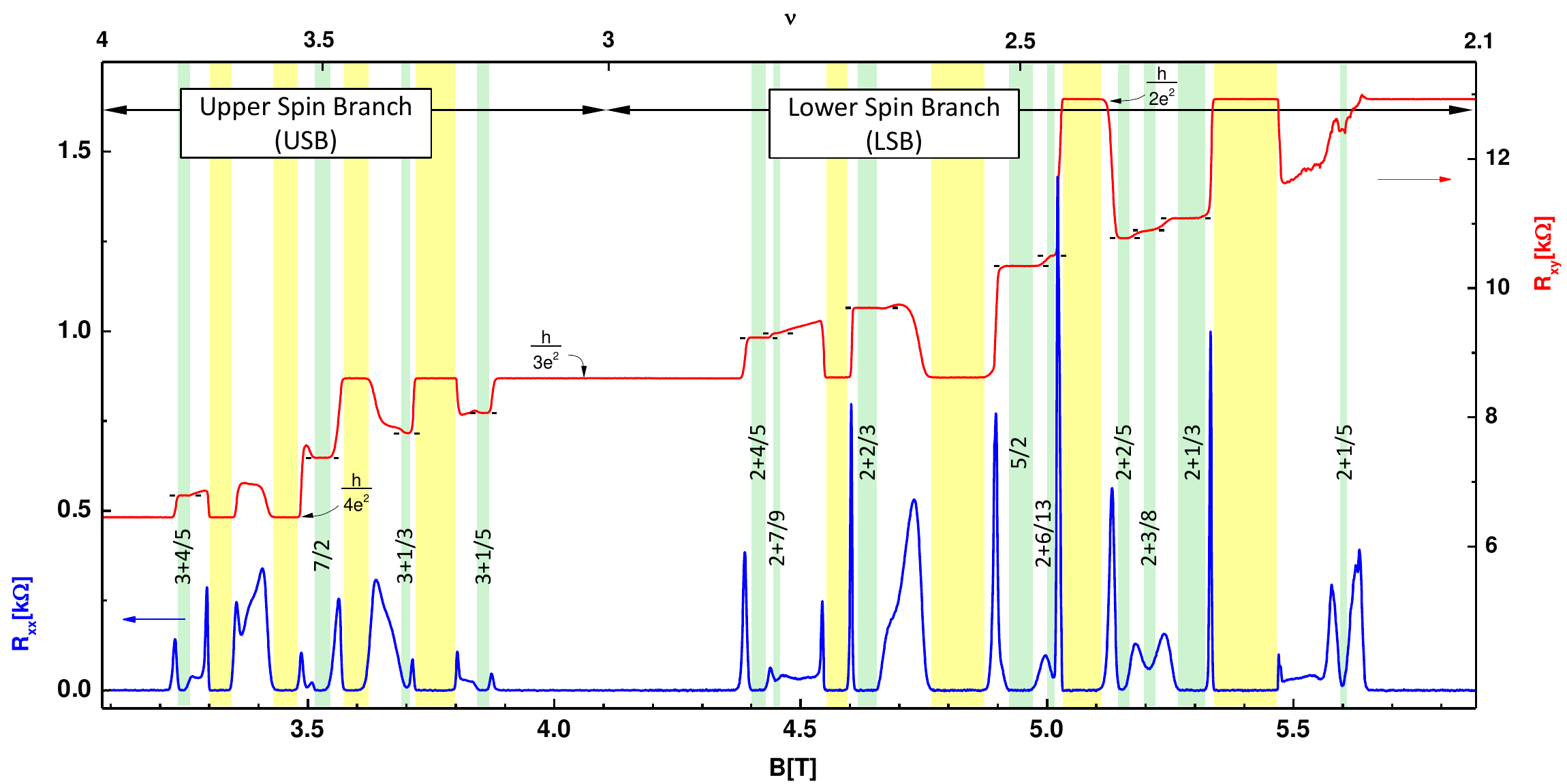}
 \caption{ Magnetoresistance traces in the second Landau level, i.e. in the filling factor range $2<\nu<4$,
 measured at $T=6.9$~mK.
The region of the lower spin branch (LSB) and upper spin branch (USB) are clearly marked.
Fractional quantum Hall states are shaded in green, while the reentrant integer quantum Hall states in yellow.
% Data in the lower spin branch is from Ref. \cite{S-Kumar}. 
The overall symmetry between these two spin braches is 
evident by the development of fractional quantum Hall states and reentrant integer quantum Hall states
at similar partial filling factors. 
 \label{Fig1}}
 \end{figure*} 
 
 In order to thermalize electrons close to their ground state
we utilize our ultra-low temperature setup consisting of a He-3 immersion cell \cite{pan99,Samkharadze1}.
Cooling is ensured by eight sintered silver heat exchangers are
immersed in the liquid He-3 bath.  Thermometry is performed using a quartz tuning fork viscometer 
which monitors the temperature dependent viscosity of the He-3 bath \cite{Samkharadze1}.

\begin{figure}[t]
 \includegraphics[width=1.0\columnwidth]{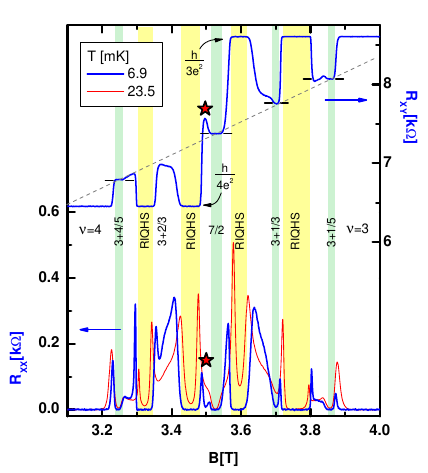}
 \caption{ The magnetoresistance in USB SLL ($3<\nu<4$) at
 6.9~mK as function of the magnetic field $B$ (bottom scale) and filling factor (upper scale).
 Numbers mark various filling factors of interest. We note the absence of a green shading at $\nu=3+2/3$,
 as a FQHS does not develop here, even though a local minimum is present in $R_{xx}$
 at this filling factor.
The dashed line is the classical Hall line and the star symbol is indicative of a developing
RIQHS of a new type described in the text.
 \label{Fig2}}
 \end{figure} 

We measured a high quality sample, in which we have already studied transport in
the  LSB of the SLL \cite{kumar10}.  Figure 1 shows this region of the LSB SLL 
at magnetic fields  $B > 4.1$~T. In this region  at $\nu=f$ we observe a multitude of FQHSs
as distinguished by
a vanishing longitudinal magnetoresistance $R_{xx}$ and Hall resistance $R_{xy}$ quantized to $h/f e^2$ \cite{tsui}.
We also observe four reentrant integer quatum Hall states (RIQHSs) signaled by quantization of $R_{xy}$
to an integer, either $h/2e^2$ or $h/3e^2$ \cite{eisenstein, Deng}.  

Extending measurements to lower $B$-fields, we access the USB SLL. As seen in Fig.1, in this region
we observe known FQHSs at  filling factors $\nu=7/2$, $3+1/5$, $3+4/5$ \cite{eisenstein} and
four RIQHSs \cite{eisenstein, Deng}.  These FQHSs and RIQHSs form at the same partial filling factors,
defined as decimal part of the filling factor $\nu$. The various ground states in the two spin brances
are connected by particle-hole symmetry \cite{girvin}, therefore
the ground states at $\nu$,  $5-\nu$, $1+\nu$, and $6-\nu$ are said to be symmetry-related
or conjugated states.
For example, the  FQHSs shown in Fig.1 at $\nu=2+1/5$, $2+4/5$, $3+1/5$, and $3+4/5$ 
belonging to the different spin branches are symmetry-related. 
% Thus the pairs of FQHSs at $\nu=2+1/3$, $3+1/3$ and  $2+1/5$, $3+1/5$ are symmetry-related. 

As seen in Fig.1, strong local minima in $R_{xx}$ also develop in the USB SLL at 
$\nu=3+1/3$ and $\nu=3+2/3$. However, the presence of these 
minima does not guarantee the formation of a FQHS at these filling factors. It is known that at $\nu=1/7$, for example, 
no FQHS develops even though a depression in $R_{xx}$ is present at finite temperatures \cite{seven}.
A defining feature of a FQHS, and of any topological ground state in general, 
is the opening of an energy gap in the bulk of the sample. An energy gap  $\Delta$ is signaled
by an activated  magnetoresistance $R_{xx}$ with a $T$-dependence of the form  $R_{xx}\propto e^{-\Delta/2k_B T}$.
Other hallmark properties of a FQHS are a quantized Hall resistance $R_{xy}$ and a  vanishing $R_{xx}$ in the limit of $T=0$ \cite{tsui}.
While weak indications of FQHSs have been reported at $\nu=3+1/3$ or $3+2/3$ in Ref.\cite{eisenstein},
none of the above described hallmark properties of a FQHS have been observed.
A close-up of the USB SLL is shown in Fig.2. We can see that
at $\nu=3+1/3$, our $T=6.9$~mK data exhibit both a vanishingly small $R_{xx}$ as well as
an $R_{xy}$ consistent with a plateau quantized to $h/(3+1/3)e^2$.
% to within 0.13\%. 

Magnetotransport at $\nu=3+2/3$, however, is markedly different from that at $\nu=3+1/3$.
As seen in Fig.2, $R_{xx}$ develops a local minimum at $\nu=3+2/3$. However, as seen in Fig.2,
$R_{xy}$ at $\nu=3+2/3$ clearly deviates from the quantum value $h/(3+2/3)e^2$,
the expected value for a FQHS at this filling factor,
%, despite the local minimum in $R_{xx}$ at this filling factor.  
casting a doubt on whether the ground state at $\nu=3+2/3$
is a FQHS. Furthermore, as also shown in Fig.2, $R_{xx}$ at $\nu=3+2/3$ increses with a decreasing temperature,
suggesting that $R_{xx}$ does not vanish as $T$ is lowered.

A detailed temperature dependence of the $\nu=3+1/3$ and $3+2/3$ FQHSs is shown in Fig.3b.
Demonstrated by the linear segments in the arrhenuis plots shown in Fig.3b,
$R_{xx}$ measured at $\nu=3+1/3$ is found to be activated. 
The opening of an energy gap  $\Delta_{3+1/3}=37$~mK unambiguously establishes, 
for the first time, the formation of a new FQHS at $\nu=3+1/3$. 
From data shown in Fig.3a and Fig.3b, we extract the energy gaps of 
the other odd-denominator FQHSs in the SLL:  $\Delta_{3+1/5}=104$~mK, $\Delta_{3+4/5}=113$~mK, 
$\Delta_{2+1/5}=210$~mK, and $\Delta_{2+4/5}=212$~mK. 
Errors due to scatter in the data amount to $\pm 5\%$. 

Fig.3b also reveals that the  $T$-dependence at $\nu=3+2/3$, in contrast to that at
$\nu=3+1/3$, is not activated. The  FQHS at $\nu=3+2/3$ thus
does not develop an energy gap in our sample in spite of the presence of a local minimum in $R_{xx}$.
The ground state at $\nu=3+2/3$ is therfore not a FQHS.
%   Becasue of the reduced $B$-field at which it develps, we think that it is likely that a FQHS at $\nu=3+2/3$ is abent because of disorder. 
% The negative value is because of neg term Eq.1.
However, the  emergence of a fractional quantum Hall ground state at this filling factor
in future higher quality samples cannot be ruled out at this time.

\begin{figure}[b]
\includegraphics[width=0.95\columnwidth]{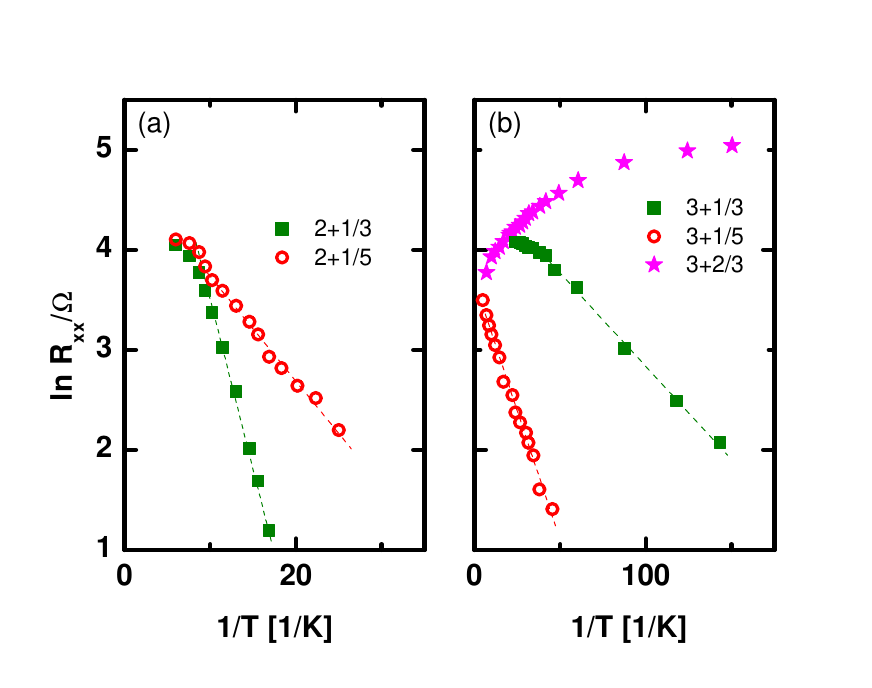}
\caption{Arrhenius plots of the $R_{xx}$ minima at several odd-denominator filling factors in the 
LSB (panel a.) and USB (panel b.) of the SLL.
%(a) $\nu=2+1/5$ and $2+1/3$ from the LSB and (b) $\nu=3+1/5$, $3+1/3$, $3+2/3$ from the USB. 
Data at $\nu=2+1/3$ is from Ref.\cite{kumar10}.
% Dotted lines are linear fits and show the range of activated transport.
 \label{Fig3}}
\end{figure}

%In the limit of large magnetic fields $B$ ground states within different spin branches 
%are similar and are linked via particle-hole symmetry. 

% Gaps of symmetry-related FQHSs are expected to scale the same way.
%are symmetry related. Aspects of particle-hole symmetry 
%are well documented in the LLL \cite{Yeh}[pan] and have been addressed
% and in the limit where Landau level mixing can be neglected,  in the SLL for the $\nu=5/2$ FQHS
%as well as for the RIQHSs \cite{Deng,eisenstein,Shayegan1,Shayegan2}.
%
%Experiments so far did not detect any sign of a spin transition in either the $\nu=2+1/3$ or the $2+1/5$
%FQHS. A spin transition has been observed in the $\nu=2+2/3$ FQHS, albeit at the 
%low value of $B$=1.XX~T. 

The energy gaps in the LSB SLL satisfy the $\Delta_{2+1/3}>\Delta_{2+1/5}$ relationship. This is
a well established inequality in many samples of various electron densities
 \cite{pan99,dean08,choi08,kumar10,nuebler10}.
Similar inequalities are also known in the lowest Landau level. Indeed,
$\Delta_{1/3}>\Delta_{1/5}$ found in the LSB LLL \cite{Willett3,Mallett,vonKlitzing}.
Furthermore,  there is evidence that in the USB LLL the $\nu=1+1/3$ FQHS is more prominent than 
the $\nu=1+1/5$ FQHS \cite{Eisenstein1,Sachrajda}. Therefore it appears that
the FQHS at partial filling factor 1/3 is more stable (i.e. it has a larger energy gap) than that
at partial filling 1/5. To our surprise, however, this generally observed relationship is reversed
in the USB SLL. Specifically, we find that $\Delta_{3+1/5}>\Delta_{3+1/3}$.
This anomalous gap reversal in the USB SLL indicates an
unanticipated difference between the prominent odd-denominator FQHSs forming in different spin branches.

The anomalous $\Delta_{3+1/5}>\Delta_{3+1/3}$ gap reversal may be caused 
by a suppression of the FQHS at $\nu=3+1/3$ due to a spin transition in this state.
% Indeed, within the CF model, at $\nu=3+2/3$ one has a single CF $\Lambda$-level filled with holes. 
Experiments so far have not detected any sign of a spin transition in either the $\nu=2+1/3$ or the $2+1/5$ FQHSs
and NMR measurements at $\nu=2+1/3$ are consistent with fully spin polarizated state \cite{nuebler10,muraki,panSpin}.
While a spin transition has recently been observed in a related FQHS at $\nu=2+2/3$ \cite{panSpin}, this transition
occurs at a magnetic field $B \sim 1.24$~T considerably lower than the field $B=3.7$~T 
the $\nu=3+1/3$ FQHS forms in our sample.  
%within the Laughlin-Jain model, the $\nu=3+1/3$ is fully spin polarized,
%therefore a spin transition is not expected.
We thus think spin is not likely to play a significant role in a possible suppression of 
a FQHS at $\nu=3+1/3$. 

An anomalous $\Delta_{3+1/5}>\Delta_{3+1/3}$ gap reversal may also be caused by Landau level mixing \cite{yoshioka} or finite width
effects \cite{PetersonJolicoeur}. Landau level mixing is a gap reducing effect due to the unoccupied Landau levels above the Fermi energy 
and its magnitude is enhanced with reduced $B$-fields. Similarly, finite width effects change with the $B$-field 
as they scale with the $w/l_B$ ratio. Here $w$ is the width of the quantum well and $l_B=\sqrt{h/eB}$ the magnetic length.
In our sample the $\nu=3+1/3$ and $3+1/5$ FQHSs develop at lower magnetic fields than
their symmetry related counterpart FQHSs at $\nu=2+1/3$ and $2+1/5$ and may be 
infuenced by the two effects discussed above. 
However, a reversal of the $\Delta_{2+1/3}>\Delta_{2+1/5}$ inequality has never been 
detected in any experiments, even when the electron densities are as low as
$n \approx 7.7 \times 10^{10}$/cm$^2$  \cite{nuebler10,panSpin}.
%                 We conclude that a gap reversal has never been seen for the  $\nu=2+1/3$ and $2+1/5$ FQHSs.
We thus conclude that the gap reversal of the prominent odd-denominator FQHSs of the SLL
is not present in the LSB at any sample conditions, therefore it is an exclusive characteristic of the USB.

With spin and Landau level mixing effects ruled out, we find that the anomalous gap reversal of the
$\nu=3+1/3$ and $3+1/5$ FQHSs cannot be accounted for within the Laughlin-Jain description.
One possible cause for this anomalous behavior is the formation of fundamentally different FQHSs in different spin branches.
An alternative possibility is that FQHSs at the sample partial fillling are the same in different spin branches, but
there are fundamental differences between the states with partial filling 1/3 and those with 1/5.
It is interesting to note that this latter scenario is supported by the results of a recent experiment in which
the second electrical subband of a quantum well was populated \cite{liu11}. 
In this experiment, populating the second subband had qualitatively different effects on the
FQHSs at partial filling 1/3 and 1/5 in the LSB SLL.It was found that the $2+1/3$ and $2+2/3$
FQHSs became more robust, whereas the $2+1/5$ and $2+4/5$ FQHSs were destroyed \cite{liu11}.
The anomalous gaps we found and the contrasting results  reported in Ref.\cite{liu11}
highlight the lacunar understanding of the prominent odd-denominator FQHSs of the SLL and 
even elicits the provocative possibility that
some of the FQHSs  may not be a conventional Laughlin-Jain type, but 
rather of an unknown origin. We note that the possibility that the odd-denominator FQHSs studied here
are of non-Laughlin type cannot be ruled out on theoretical grounds. Indeed, as mentioned in the introduction,
the overlap of the exact and numerically obtained wavefunctions is not satisfactory for a firm assignment 
of these states to Laughlin states \cite{wojs2,Toke,Papic,PetersonJolicoeur,SimionQuinn,wojs1}
and alternative theories exist which are distinct from the Laughlin-Jain construction \cite{ReadRezayi,Ardonne,Simon,BondersonSlingerland,WojsJain}.

At a given filling factor the theory allows the existence of 
several fundamentally different ground states and even considers
transitions induced between these different states.
At the root cause of the formation of the various ground states
we find minute differences in 
the effective electron-electron interaction potential   
caused by changes in the sample parameters. 
% The FQHSs in the SLL are especially sensitive to small changes in this effective interaction.
We think that the anomalous gap reversal observed is due to the difference in the 
effective electron-electron interaction when we populate either  the USB SLL or the LSB SLL.
We surmise that the study of FQHSs at sample parameters which modify the electron-electron interactions
may therefore be of fundamental importance in tuning topological order
and may provide a pathway to discovery of novel topological ground states. 

Our data reveal that the modified electron-electron interations in the SLL have another 
unforeseen consequence. As seen in Fig.2,  at the location $B=3.50$~T of the star symbol,
$R_{xx}$ is nearly vanishing and  $R_{xy}$ exceeds the classical Hall value. Such a behavior is inconsistent with a FQHS;
we think it is a signature of an incipient RIQHS. However, 
this incipient RIQHS observed at $B=3.50$~T, is different from the known RIQHSs \cite{eisenstein,Deng}.
Indeed, the two known RIQHSs at $\nu>7/2$, which develop at $B=3.32$~T and $3.45$~T
have $R_{xy}$ qunatized to $h/4e^2$, whereas the incipient RIQHS at $B=3.50$~T
appears to develop towards $h/3e^2$  in the limit of $T=0$.   
% A RIQHS with such unusual properties
% is not observed in the LSB SLL at a $B$-field just above that of $\nu=5/2$ (see Fig.1S of the Supplement).

In summary, we have probed the upper spin branch of the second Landau level which 
appears to be richer than originally thought. Our energy gap measurements 
of the odd-denominator FQHSs in this region 
allowed for a test of the symmetry relations between these FQHSs and
revealed an unexplained relative magnitudes of these energy gaps.  
Furthermore, we observed a nascent RIQHS of an unusual type.
% which may yield valuable insight on the nature of these states and stimulate further theory work.

Measurements at Purdue were funded by the NSF grant DMR-1207375 and the sample growth efforts of L.N.P. and K.W.W. 
were supported by the Princeton NSF-MRSEC and the Moore Foundation. We thank  N. d'Ambrumenil
and Z. Papi\'{c} for useful discussions.

%We measured a 4$\times$4~mm$^2$ piece of a 30~nm wide GaAs/AlGaAs quantum well sample in a van der Pauw geometry. 
%The density and mobility are $n=3.0\times 10^{11}\: \mbox{cm}^{-2}$ and $\mu=32\times 10^6\: \mbox{cm}^{2}/\mbox{Vs}$,
%respectively. This sample has the advantage that the $\nu=2+1/3$ and $2+2/3$ FQHSs
%from LSB SLL have previously been studied \cite{S-Kumar} and produced record values of energy gaps for several FQHS in that region.

%\begin{figure}
%{\centering
%\includegraphics[width=.5\textwidth]{Gaps2.pdf}
%}
%\caption{Arrhenius plots of the $R_{xx}$ minima at several odd-denominator filling factors in the SLL such as
%(a) $\nu=2+1/5$ and $2+1/3$ from the LSB and (b) $\nu=3+1/5$, $3+1/3$, $3+2/3$ from the USB. Data at $\nu=2+1/3$
%is from Ref.\cite{Kumar}. Dotted lines are linear fits and show the range of activated transport.}
%\end{figure}

%\begin{thebibliography}{l}
%\bibitem{S-Kumar} A. Kumar, G.A. Cs\' athy, M.J. Manfra, L.N. Pfeiffer, and K.W. West, Phys. Rev. Lett. \textbf{105}, 246808 (2010).
%\end{thebibliography}

\end{document}